# Electrodynamic Response of MgB$_2$ Sintered Pellets and Thin Films

A. Andreone, G. Ausanio, E. Di Gennaro, G. Lamura, and M. Salluzzo
I.N.F.M. and Dipartimento Scienze Fisiche, Universita' *Federico II*, Napoli, Italy

J. Le Cochec
Laboratoire Pierre Süe, DRECAM, CEA-Saclay, 91191 Gif sur Yvette, France

A. Gauzzi
MASPEC-CNR Institute, Parma-Fontanini, Italy

C. Cantoni and M. Paranthaman
Oak Ridge National Laboratory, Oak Ridge, Tennessee, U.S.A.

G. Giunchi and S. Ceresara
Edison S.p.A., Milano, Italy

## I Introduction

The discovery of superconductivity in hexagonal MgB$_2$ [1] with transition temperature $T_c$ near 40 K has led to considerable interest in this material, since it offers a new class of simple and low-cost binary intermetallic superconductors with a record high $T_c$ for non-oxide and non C-60 based compounds. The observed critical temperature seems to be located either close to or just beyond the theoretical upper limit predicted for phonon-mediated superconductivity [2], almost a factor two larger than previously known intermetallic superconductors and comparable to the value found in some high-$T_c$ superconductors. An impressive number of theoretical and experimental studies has been performed on this new superconductor to inspect fundamental issues, like the nature of the pairing mechanism and the origin of superconductivity. Mostly, experimental results have provided evidence for a phonon mediated superconductivity. The boron-isotope effect [3], the observation of a tiny coherence peak just below the critical temperature $T_c$, and the spin lattice relaxation rate in the superconducting state in nuclear magnetic resonance (NMR) [4], Raman spectroscopy [5], tunneling measurements [6-8] and optical conductivity data [9] are all consistent with the existence of a conventional mechanism of superconductivity. The effect of impurities has been also investigated: chemical substitution of magnetic ions such as $Mn^{2+}$, $Fe^{2+}$, $Co^{2+}$ and $Ni^{2+}$ suppresses the critical temperature $T_c$ [10] while non magnetic Zn-substitution increases it [11]. Although it seems that the body of the experimental data can be explained in a BCS-conventional framework, no consensus has been reached on the pairing symmetry of this system yet. The majority of spectroscopic measurements have shown so far a finite but widely variable value for the energy gap, ranging from 2 to 7 meV. At present, the most plausible explanation is being considered the presence in $MgB_2$ of two different superconducting energy gaps arising from two different sheets of the Fermi surface [12].

As far as applications are concerned, the wide availability of a new compound exhibiting a superconducting phase within the easy reach of light, compact, fast cryocoolers has prompted also a discussion on whether $MgB_2$ might have an impact on the future viability of superconducting electronics. Moreover, its high metallic nature as well as the strongly linked character of grain boundaries in polycrystalline samples [13] makes $MgB_2$ a promising candidate for the development of microwave passive devices for wireless communications [14] and superconducting cavities for particle accelerators [15].

In this context, the study of the electrodynamic response of this material offers a twofold interest:
- measurements of the magnetic penetration depth and microwave surface resistance can help in shedding a light on the nature of superconductivity in diborides, since they both probe the low-lying quasi-particle excitations;
- the investigation of microwave losses is directly related to technological issues of outmost importance, like the power handling capability of devices and the field dependence of accelerating cavities.

In this paper we will summarize a number of experimental results on the temperature dependence of the magnetic penetration depth $\lambda$ and on the temperature and field dependence of the microwave surface impedance $Z_s = R_s + iX_s$ in both pellets and thin films of $MgB_2$, exhibiting critical temperatures ranging between 26 and 38 K. The measurements of $Z_s(H,T)$ were performed by means of a sapphire dielectrically loaded cavity operating at 20 GHz, whereas the study of $\lambda(T)$ was carried out employing a single coil mutual inductance technique in the MHz region.

## II Samples preparation and characterization

Two high density sintered pellets (#1A and #2A) and two films (#1B and #2B) deposited on a r-plane single-crystal sapphire substrate have been used in this study.

Bulk samples were obtained by reaction sintering of elemental B and Mg for 3 h at 950 °C in a sealed stainless steel container, lined by a Nb foil. Details on this preparation technique are described elsewhere [16, 17]. The resulting high density (~ 2.4 g/cm$^3$) $MgB_2$ cylinders had metallic appearance and very high hardness. Two samples (10x10x3 mm$^3$) were cut from the bulk material with a diamond circular saw.

The mechanically polished surface of these samples was observed by optical microscopy revealing the presence of large single crystal grains (up to 200 μm) [13, 18-20]. Those grains are composed of fine-grained hexagonal MgB$_2$ crystallites (between 100 to 500 nm wide) as revealed by atomic force microscopy (AFM). Figures 1(a) and 1(b) show amplitude images of the samples surface obtained using a commercial Nanoscope III (Digital Instruments) AFM operated in contact mode.

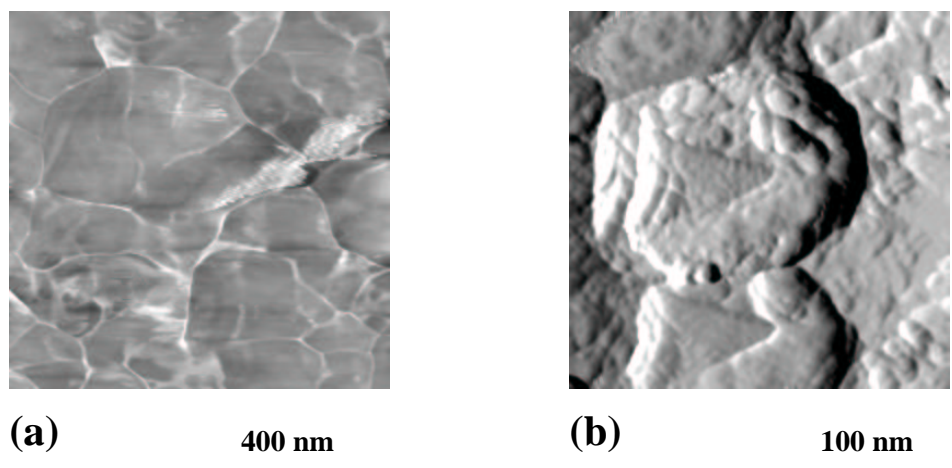

(a) 400 nm  (b) 100 nm

**Figure 1:** AFM scan amplitude in contact mode on the surface of sample #2A as cast (a) and mechanically polished (b).

To reduce the presence of an overlayer of oxidized Mg [21], pellets are immersed for 60 s in a non-aqueous etchant (1% HCl in absolute ethanol), rinsed in ethanol and blown dry with nitrogen immediately before measurements.

Films were obtained by appropriate postannealing of an electron-beam-evaporated B precursor (500 nm) grown directly on a r-plane Al$_2$O$_3$ single-crystal substrate. After deposition the boron film was subsequently sandwiched between cold pressed MgB$_2$ pellets, along with excess Mg turnings, packed inside a crimped Ta cylinder and finally annealed at 890 °C for 20-25 min [22]. Typical X-ray diffraction θ-2θ scan shows that samples have the c-axis perpendicular to the surface, whereas the pole figure evidences a random in-plane texture. In figure 2(a) and its magnified view (fig. 2(b)) the AFM images taken on one of these films (# 1B) are shown. The surface is composed of well connected fine grains having an average dimension of 200 nm and a fairly high roughness of 80 nm r.m.s. Such a high value has to be related to the *ex-situ* synthesis process: AFM performed on *in situ* grown films indicates a much smoother surface (15 nm r.m.s.) for a comparable area (500 nm x 500 nm) [23]. In figure 3(a) and its magnified view (fig. 3(b)) the same sample is shown after removing a 60 nm thick layer by an ion (Ar$^+$) beam etching. Grains on the surface appears now to have an hexagonal-like shape which can be related to the in-plane lattice. AFM analysis shows that ion milling reduces surface roughness to 70 nm r.m.s.

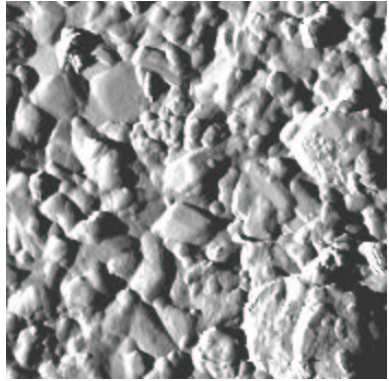 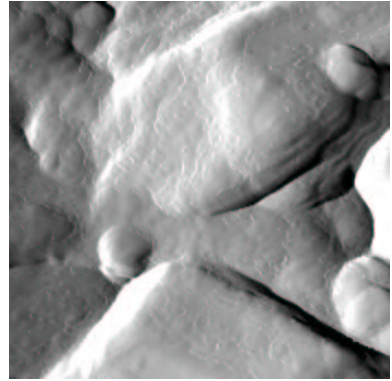

(a) 1 μm  (b) 400 nm

**Figure 2:** AFM amplitude images in contact mode on the surface of sample #1B: (a) large scan area; (b) details of the grains.

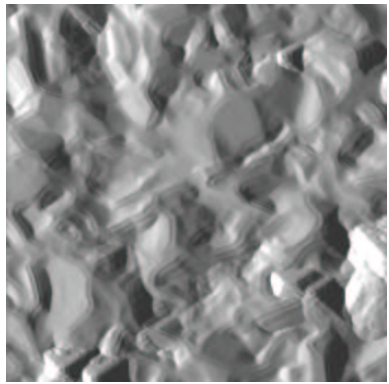 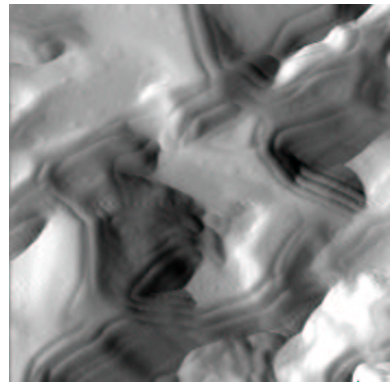

(a) 1 μm  (b) 300 nm

**Figure 3:** AFM amplitude images in contact mode on the surface of sample #1B after ion milling: (a) large scan area; (b) details of the well-connected grains. Note the hexagonal shape of the crystallites.

In table I the main superconducting and transport parameters for the samples under study are summarized. The critical temperature $T_c$ and critical current density $J_c$(4 K) values were inductively measured. The room temperature resistivity was estimated using a standard four point probe technique. $\lambda(0)$ was determined through a fit of the temperature dependence of $\Delta\lambda$, as shown in the next section. The residual surface resistance $R_{res} = R_s$ (4

K) values is also reported. In the case of films, the $R_s$ values have been corrected for the finite thickness [24].

| Sample | $T_c$ (K) | $\Delta T_c$ (K) | $J_c$ @ 4 K (MA/cm$^2$) | $\rho_{300}$ ($\mu\Omega\cdot$cm) | $\lambda(0)$ ($\mu$m) | $R_{res}$ (m$\Omega$) |
|---|---|---|---|---|---|---|
| Pellet #1A | 38.4 | 1.3 | — | 6 | 2.4 | 5.6 |
| Pellet #2A | 37.0 | 2.3 | — | — | 1.6 | 6.8 |
| Film #1B | 26.0 | 1.0 | 0.4 | 900 | 1.3 | 0.4 |
| Film #2B | 37.9 | 1.0 | 6.4 | 22 | 0.1 | 1.7 |

**Table I:** the main transport parameters measured for the samples under study.

### III Experimental techniques

**III.1 Radiofrequency magnetic penetration depth measurements**

One of the most reliable techniques employed for investigating the nature of pairing in superconductors is the measurement of the magnetic penetration depth $\lambda$, which is known to be a very sensitive probe of the low-lying quasiparticles states. Since the information provided by penetration depth data are significant on a $\lambda(0)$ scale rather than on a $\xi(0)$ scale, for a London superconductor one can safely assume that surface quality would affect the results much less than it does in tunneling data. Moreover, in the case of a thin film this would correspond to probe the true "bulk" properties of the superconductor.

First experiments performed on MgB$_2$ pellets using muon spin resonance ($\mu$SR) [25] and ac susceptibility [25, 26] reported a quadratic behavior of $\Delta\lambda$ at low temperatures, that the authors interpreted as the signature of the presence of nodes in the gap. Very recently, however, a more careful analysis of $\mu$SR data showed [27] results consistent with a two-gap model, while the peculiar exponential behavior expected for a s-wave superconductor was indeed observed in polycrystalline samples [28], in wires [29], and recently in single crystals [30]. Measurements carried out on films coming from different sources show also a fully-gapped superconductor behavior, even if quite different values of the strong coupling ratio are reported [31, 32].

In the following the temperature dependence of the magnetic penetration depth, $\lambda(T)$, at low frequency (4 MHz) is studied, using a single coil-mutual inductance technique [33]. This technique measures the change of inductance $\Delta\lambda$ of a pancake coil located in the proximity of the sample and connected in parallel with a low-loss capacitor. Such LC-circuit is connected to the input of a marginal oscillator. A change of impedance of the LC-circuit is detected as a change of resonant frequency $f$ of the oscillating signal:

$$f = \frac{1}{2\pi}\sqrt{\frac{1}{L(T)C(4K)} - \frac{R(T)^2}{L(T)^2}} \qquad (1)$$

where L is the total inductance of the system formed by the coil and superconducting sample, C is the capacitor and R a function of the coil resistance and of the sample surface losses. In the Meissner regime, where the electromagnetic response of the sample is mostly diamagnetic and the losses are negligible, the explicit expression of the variation of the coil inductance as a function of the magnetic penetration depth is:

$$\Delta L = \pi \cdot \mu_0 \int_0^\infty \frac{M(\gamma)d\gamma}{1 + 2\cdot\gamma\cdot\lambda\cdot\coth\left(\dfrac{d}{\lambda}\right)} \qquad (2)$$

where $M(\gamma)$ is a geometrical factor depending on the sample-coil distance and $d$ is the sample thickness. The details of this technique are described elsewhere [33].

### III.2 Microwave magnetic penetration depth measurements

Measurements of the surface impedance $Z_s(T, H) = R_s(T, H) + iX_s(T, H) = R_s(T, H) + i\omega\mu_0\lambda(T, H)$ in the microwave region ($\omega$ is the angular frequency) were carried out by using a dielectric resonant cavity technique. The resonator consists of a cylindrical cavity of diameter D = 9.5 mm short-circuited at both ends by two plates. A cylindrical dielectric sapphire rod of diameter d = 7 mm and height h = 3.5 mm is placed and centered between the two parallel plates (see fig. 4). The $TE_{011}$ mode of the resonator is excited and detected by two semi-rigid coaxial cables, each having a small loop at the end, and the resonant frequency $f$ and Q-factor are measured in the transmission configuration.

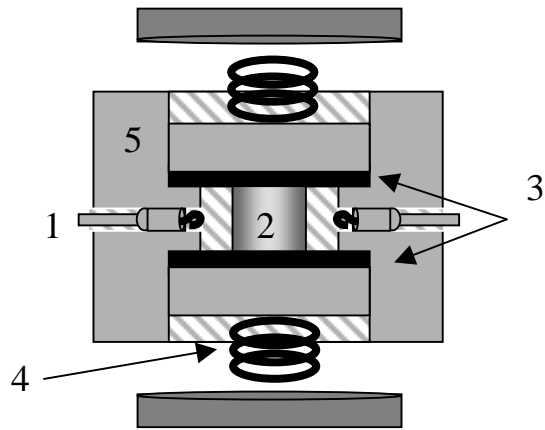

**Figure 4:** schematic drawing (not in scale) of the dielectrically loaded resonator operating at 20 GHz for $Z_s(T, H)$ measurements: coupling ports (1), sapphire puck (2), superconducting sample(s) under test (3), Cu-Be springs (4), cavity walls (5).

For the measurements of the surface impedance as a function of the temperature, an OFHC (Oxygen Free High Conductivity) copper cylindrical cavity is employed, where the upper and lower plates are substituted by two pellets or films having nominally the same structural, transport and superconducting properties. It is evident that this "symmetric" configuration, would the samples under test be inhomogeneous, tests only the average electrodynamic response of the material.

In the case of films, the effective (measured) surface impedance $Z_{seff} = R_{seff} + j\mu_0\omega\lambda_{eff}$ is related to the surface impedance $Z_s$ of the "bulk" material in the full temperature range but very close to $T_c$ through the equation [24]:

$$Z_{s_{eff}} = R_s \left[ \coth(t/\lambda) + \frac{t}{\lambda} \frac{1}{\sinh^2(t/\lambda)} \right] + j\mu_0\omega\lambda \coth(t/\lambda) \tag{3}$$

where t is the sample thickness.

The effective surface resistance $R_{seff}$ is obtained measuring the quality factor of the resonator in unloaded condition $Q_u$, neglecting radiation and dielectric contribution, and using the formula:

$$R_{s_{eff}}(T) = \frac{\Gamma}{Q_u(T)} - \frac{\Gamma^2}{\Gamma + \Gamma_{lat}} \frac{1}{Q_{Cu}(T)} \tag{4}$$

Here $\Gamma$ is the geometrical factor associated with the field distribution on the film surface, whereas $\Gamma_{lat}$ takes into account contribution to losses due to the lateral copper walls. $Q_{Cu}$ is the unloaded quality factor of the sapphire resonator measured replacing the superconducting endplates with two OFHC copper foils. Below 50 K and at about 20 GHz, the surface impedance of copper is well described by the anomalous skin effect regime, therefore $Q_{Cu}$ can be considered as independent of temperature. For this geometry it is $\Gamma \ll \Gamma_{lat}$, so that the correction due to the lateral walls is within 15% in the overall measurement range.

The effective change in penetration depth $\Delta\lambda_{eff}$ is extracted using the relation:

$$\Delta\lambda_{eff}(T) \approx \beta(T) \left[ \left( \frac{\Delta f(T)}{f(T_{min})} \right)_{Cu} - \left( \frac{\Delta f(T)}{f(T_{min})} \right)_{Sc} \right] + \Delta\delta_{Cu}(T) \tag{5}$$

where $\Delta f(T) = f(T) - f(0) \approx f(T) - f(T_{min})$ represents the change of the resonance frequency in the unperturbed (Cu) and perturbed (Sc) configuration, $\Delta\delta_{Cu}(T)$ represents the temperature variation of the copper skin depth, and $\beta(T) = \Gamma/[\pi\mu_0 f_{Sc}(T)]$.

Here the anomalous skin effect behavior implies $\Delta f_{Cu}(T), \Delta\delta_{Cu} \approx 0$ at all temperatures of interest, so that:

$$\Delta\lambda_{eff}(T) \approx \frac{\Gamma}{\pi\mu_0 f(T_{min})} \left[ \left( \frac{f(T_{min})}{f(T)} \right)_{Sc} - 1 \right]. \tag{6}$$

For the field measurements of the surface impedance a high purity (RRR > 500) niobium cavity instead of copper was employed, taken at liquid helium temperature and replacing one end with the $MgB_2$ sample under test.

The advantages of using superconducting niobium are:

  *i*) the surface properties of the sample under test are easily determined, since one can safely assume that the niobium contribution to losses is negligible in such a configuration (D/d < 1.5);

  *ii*) the very high values of the quality factor (> $10^5$) allows to increase the power circulating in the cavity and therefore the maximum magnetic field generated on the $MgB_2$ sample surface.

The main drawback of this configuration is that the surface impedance and its field dependence can be studied only at 4 K, due to the low critical temperature of niobium. Varying the microwave power feeding the cavity, one can increase the maximum microwave magnetic field on the sample surface, $H_{max}$. This quantity determines the maximum surface current density on the superconducting plate. In a resonator of infinite diameter, $H_{max}$ is given by the formula [34]:

$$H_{max} = J_{1,max} \left\{ \frac{2\pi R_s}{P_0} \int_0^d J_1^2(\xi_1 r) r dr \left[ 1 + \varepsilon_r R + \frac{240\pi^2 \varepsilon_r \tan\delta}{R_s} \left( \frac{h}{\lambda_0} \right)^3 \right] \right\}^{-1/2} \quad (7)$$

where $J_{1,max}$ is the maximum value of the Bessel function of the first kind $J_1(\xi_1 r)$, $P_0$ is the total dissipated power in the dielectric resonator, $R$ is the ratio of the electrical energy stored outside and inside the dielectric rod, $\lambda_0$ is the wavelength in free space, $tan\delta$ and $\varepsilon_r$ are the loss tangent and the dielectric constant of the single-crystal sapphire respectively.

The maximum in the radial component of the magnetic field is reached in the cavity around a circle of radius of about 2 mm. The application of formula (7) to our case (D/d < 1.5) gives a relative error in the evaluation of $H_{max}$ below 5% [35].

### IV Magnetic penetration depth: experimental results and discussion

In figs. 5 and 6 the change of the magnetic penetration depth $\Delta\lambda(T) = \lambda(T) - \lambda(T_{min})$ is shown for all the samples investigated in a temperature range from 4 K to $T_c$. $T_{min}$ represents the lowest temperature that has been reached during each measurement. Comparing the curves, one can see that a much larger variation of $\Delta\lambda$ is observed in the case of pellets (fig. 5) with respect to films (fig. 6).

At low temperatures the difference in the behavior of the magnetic penetration depth is even more evident: film #2B displays a flat dependence (up to 6 K), which is less pronounced in film #1B (see fig. 7), whereas pellets show no evidence of saturation (see fig. 8). A simple BCS conventional s-wave calculation fits very well the low temperature data of film #2B with the following parameters: $\Delta(0)$ = 3.6 meV and $\lambda(0)$ = 120 nm, where $\Delta(0)$ and $\lambda(0)$ are the energy gap and the magnetic penetration depth at zero temperature respectively.

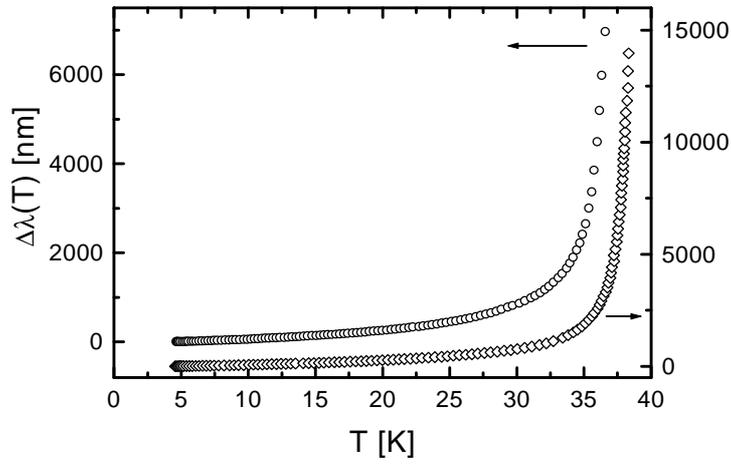

**Figure 5:** variation of the magnetic penetration depth as a function of temperature for samples #1A (○) and #2A (◊).

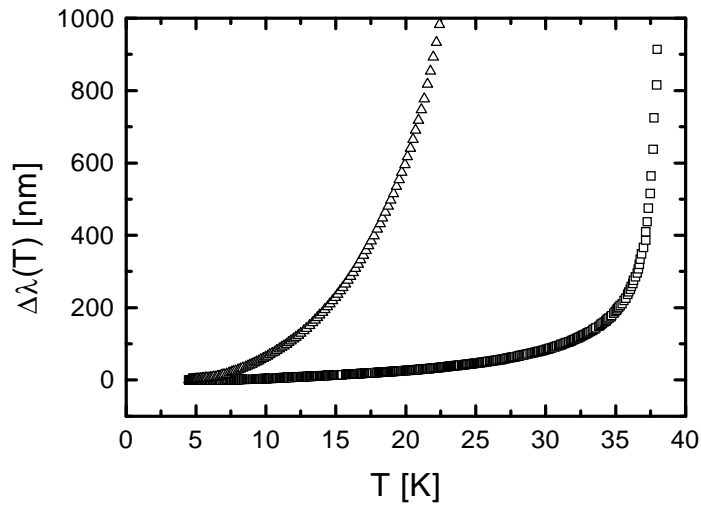

**Figure 6:** variation of the magnetic penetration depth as a function of temperature for samples #1B (△) and #2B (□).

The estimated error on the fitting parameters is about 20%. λ(0) is of the same order of magnitude of values reported by previous measurements on bulk samples [25, 28] and it is nearly half of the value obtained using a far infrared reflectivity (FIR) analysis on thin films grown by the same technique [36]. The gap value is slightly larger than the one obtained by FIR measurements [36] and it is similar to what determined by tunneling measurements [6-8].

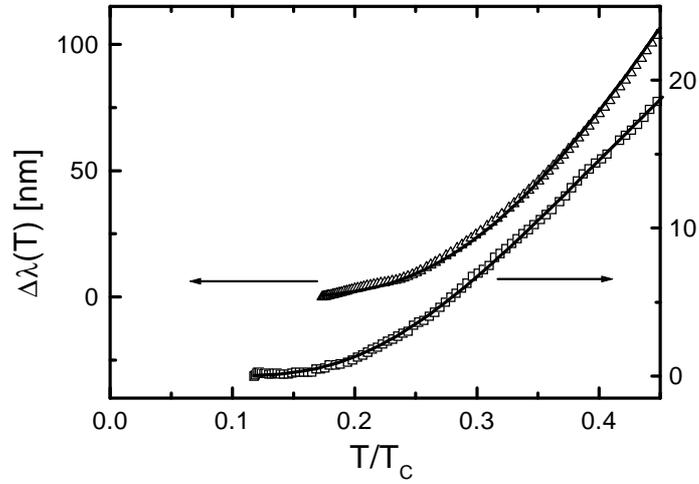

**Figure 7:** variation of the magnetic penetration depth at low temperature for samples #1B (△) and #2B (□). The solid lines represent the BCS fits.

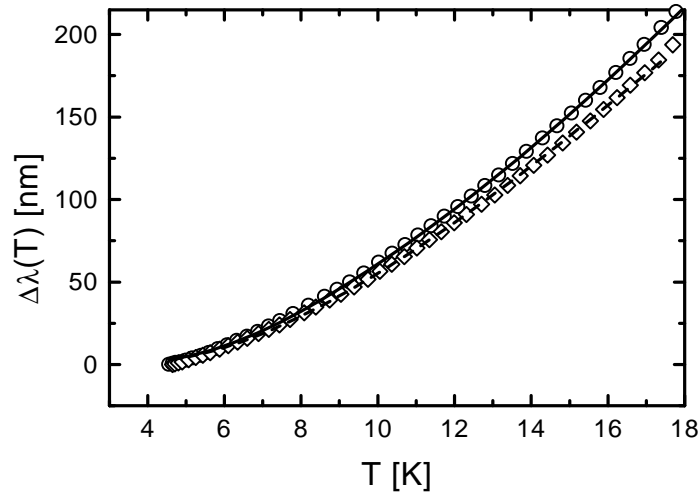

**Figure 8:** variation of the magnetic penetration depth at low temperature for samples #1A (○) and #2A (◊). The continuous and short dashed curves represent the two-gap BCS fit for #1A and #2A respectively (see text for details).

It is important to mention that the penetration depth behaviour remains the same removing by ion milling a (nominal) layer of 60 nm from the sample surface. This result rules out the proximity effect eventually associated to the presence of a metallic Mg overlayer. Therefore, if a normal or poorly superconducting layer exists on the film surface, its effect on the temperature dependence of the penetration depth is negligible in our measurements. Consequently the rather low value of the energy gap determined in our film seems, to our understanding, intrinsic to $MgB_2$. Since in our experiment the ab plane penetration depth of a c-axis oriented film is probed, this finding can be well explained by the anisotropy of the order parameter in this material reported by other studies, like point-contact spectroscopy [37], and penetration depth measurements in single crystals [30]. The result is also in agreement with the recent observation [29] that the removal of magnesium by chemical etching in a $MgB_2$ wire results in an exponential temperature dependence of the magnetic penetration depth with a similar value of the energy gap.

In the case of film #1B a BCS fit is still possible, yielding similar values of the energy gap ($\Delta(0) = 3.4$ meV) but a much larger zero temperature penetration depth ($\lambda(0) = 1.3$ μm).

The analysis of the low temperature data in pellets is more complicated. In fig. 9(a) the change in the magnetic penetration depth $\Delta\lambda$ as a function of $T^2$ is shown for both samples. An almost quadratic behavior is observed down to the lowest temperature. For comparison, the same graph is displayed in fig. 9(b) for samples #1B and #2B, clearly evidencing that a quadratic dependence is unable to explain the low temperature data for films.

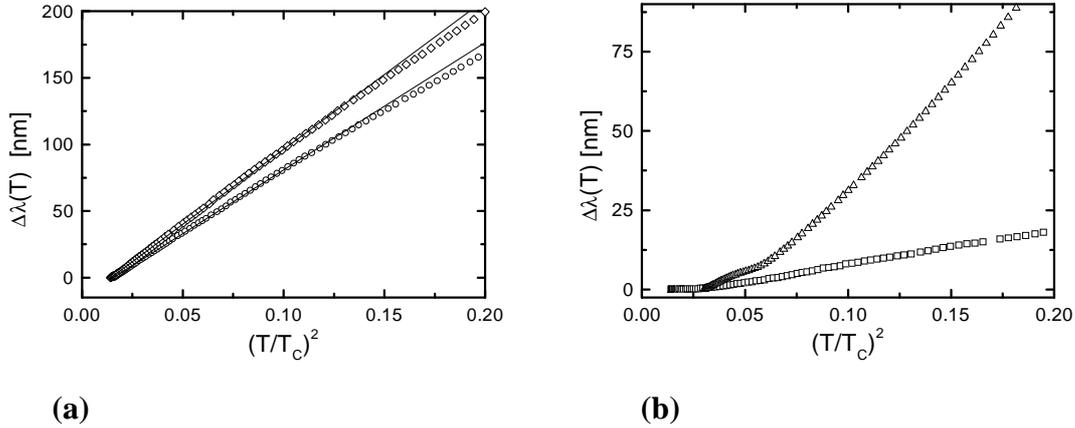

**Figure 9:** variation of the magnetic penetration depth as a function of the square of the reduced temperature (a) for samples #1A (○) and #2A (◊); (b) for samples #1B (△) and #2B (□). Solid lines are a guide to the eye

It has been claimed that the "gapless-like" behavior observed in pellets would be the signature of non conventional superconductivity in $MgB_2$ [9,25]. It is worthwhile to mention however that a $T^2$ behavior of the magnetic penetration depth is not sufficient to rule out a BCS conventional s-wave pairing symmetry. In the case of strongly inhomogeneous samples, it has been shown that a spread of superconducting gap values leads to a power-law dependence at low temperature for both the thermodynamic and transport properties [38]. Indeed the critical temperatures of samples #1A and #2A, even if grown in the same batch, differ by almost 1 K. The values of $\lambda(0)$, determined by a two fluid model fit [39], are also different: 2.4 μm for #1A and 1.6 μm for #2A respectively.

Besides that, the behavior of $\Delta\lambda(T)$ near $T_c$ is very different for the two bulk samples. These observations seem to indicate the presence of strong inhomogeneities in the pellets. In such a case a gaussian distribution of gap values in the framework of an s-wave pairing model is forcefully able to fit the experimental data.

The presence of regions of highly suppressed superconductivity, induced by unreacted Mg, can cause either a spread of gap values (from 1.7 to 7 meV as reported in literature) or a simple overlapping of two gaps [40, 41]. The magnetic penetration depth in this case provides an average of the electrodynamic response of the tested sample region. Thus it is sensitive only to the minimum and the maximum value of the gap. In the framework of a two-gap BCS s-wave model, the magnetic penetration depth can be written as a function of two energy gaps of different magnitude through the equation [42]:

$$\frac{1}{\lambda_{tot}^2} = \frac{1}{\lambda_{max}^2} + \frac{1}{\lambda_{min}^2} \tag{8}$$

where for $T < T_c/2$

$$\Delta\lambda \approx \lambda_i(0)\sqrt{\frac{\pi\Delta_i}{2k_BT}}e^{-\frac{\Delta_i}{k_BT}} \qquad (i = min, max). \tag{9}$$

Using this approach one can determine the best fit curves to the experimental data shown in fig. 8 using the following values: $\Delta_{min}^{\#1A}$=2.45 meV, $\Delta_{min}^{\#2A}$=2.4 meV, $\Delta_{max}^{\#1A}$=6.8 meV and $\Delta_{max}^{\#2A}$=6 meV respectively. The fitting parameters $\Delta_{min}$ and $\Delta_{max}$ encompass the ones found by photoemission spectra [43], tunneling spectroscopy [6, 7, 8, 40, 41, 44, 45] and specific heat measurements [46]. The corresponding strong coupling ratios $2\Delta_i/K_BT_c$ are 1.5 and 4.1 for #1A, 1.5 and 3.7 for #2A, 3.1 for #1B and 2.3 for #2B respectively.

In figure 10 results reported in the recent literature on the magnetic penetration depth temperature dependence of $MgB_2$ thin films have been collected. Data on sample #2B shows an exponential behavior similar to what observed by Klein *et al.* [31] and Zhukov *et al.* [32], in this latter case with a significantly larger value of the energy gap. For comparison, data from the μSR experiment on $MgB_2$ powders by Panagopoulos *et al.* [25], showing a quadratic temperature behavior, are also displayed.

To our understanding, the results observed on the thin film sample #2B rule out the possibility of an unconventional order parameter in $MgB_2$. A s-wave single gap model fits experimental data very well, even if with a reduced energy gap value. Although film #1B shows an increased zero temperature magnetic penetration depth and a reduced critical temperature, the experimental behavior observed at low temperatures in this sample is still compatible with what expected for a fully-gapped superconductor.

In figure 11 the normalized superfluid density $n_s$ for all samples under investigation is shown. In the case of film #1B the increase of losses increasing T prevented a reliable measurement of $\lambda$ ($\propto 1/n_s^2$) for $T > 0.9\ T_c$. For film #2B one can observe the presence of a saturation with slope zero near $T = 0$, whereas near the critical temperature the slope is -1.72, very close to the BCS expected value –2 for a conventional weakly-coupled superconductor. In the case of the pellets there is no evidence of saturation at low temperature and the slope near $T_c$ is –3.6, very close to the value expected from a simple two fluid model (–4).

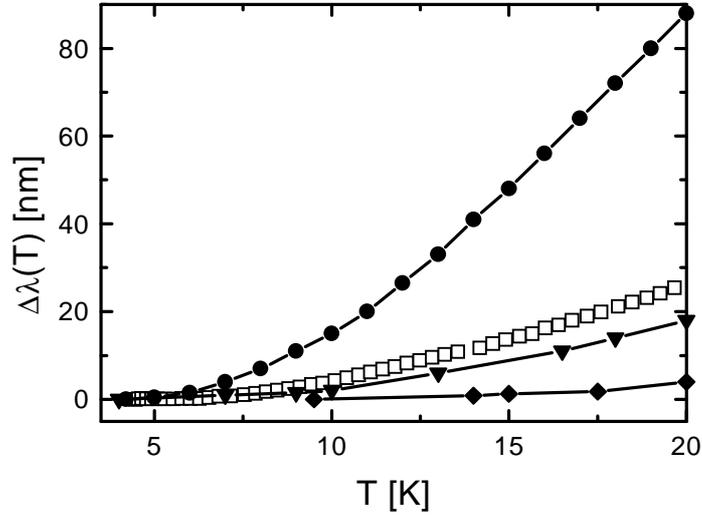

**Figure 10:** temperature shift of the magnetic penetration depth for: thin films by Klein *et al.* [31] (circles) and Zhukov et al. [32] (diamonds); sample #2B sample (open squares). Data on powders by Panagopoulos *et al.* [25] are represented by the down triangles.

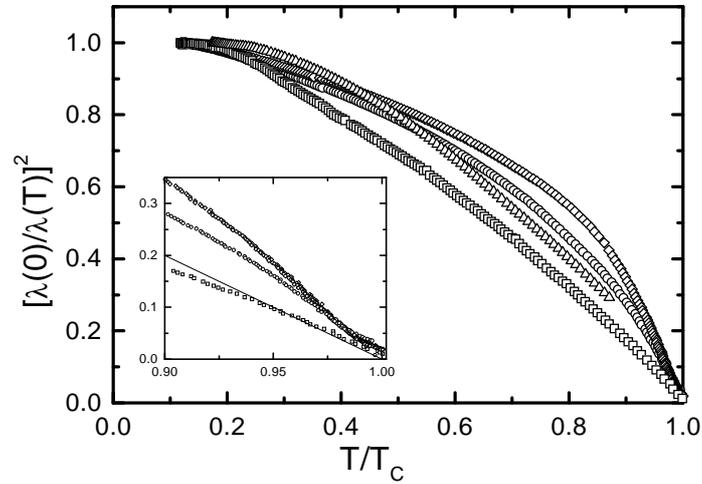

**Figure 11:** temperature variation of the normalized superfluid density for samples #1A (○), #2A (◊), #1B (△) and #2B (□). In the inset the slope expected for a BCS weak –coupled superconductor (see text).

The difference observed between films and bulk samples may have different origins. The presence of magnetic impurities cannot be ruled out since, in spite of the many research efforts spent on $MgB_2$ in the last months, the quality of available samples is not yet perfected. Thus the polynomial behavior of $\Delta\lambda(T)$ observed in pellets could be induced by magnetic impurities in a conventional s-wave superconductor [47]. Another possibility to explain the striking contrast between film #2B and pellets is to consider the anisotropy of the superconducting properties between the ab and c directions, previously mentioned, as it was recently emphasized for single crystals [48]. Since the film is c-oriented, the mutual inductance method probes mainly the ab-plane electrodynamic response, whereas this is not true for the pellets. However, the values found for $\lambda^{film}(0)$ and $\lambda^{pellet}(0)$ are not consistent with an anysotropy factor $\gamma = 2.6$ recently found in crystals [48]. Finally, the interesting coexistence of a two-dimensional Fermi surface ($p_{x-y}$ orbitals) perpendicular to the z direction and a three dimensional one ($p_z$ bonding and anti-bonding bands) reported in [12] may be of help in explaining the electrodynamic response of the pellets, which is not inconsistent with the existence of two order parameters. However, it would be difficult to explain the experimental data on thin films using the same picture. More likely, the high values of $\lambda(0)$, the difference observed in the critical temperature values and in the temperature dependence $\Delta\lambda(T)$ close to $T_c$ between the two pellets made from the same batch all address towards an explanation based on the presence of inhomogeneities: residual flakes of unreacted Mg induce a depression in the gap by the proximity effect in the region near the inclusions. In this case the presence of regions of different critical temperature depends strongly on the ratio of the mean free path to metal layer thickness and on the transmittivity properties of the N-S interface [44]. Thus the proximity effect can induce the presence of regions with a weakened gap and with a critical temperature close or even equal to the bulk critical temperature. A spread or a simple superposition of two gap values is equally possible.

## V Microwave surface impedance: experimental results and discussion

In table I the values of the residual surface resistance in the microwave region have been summarized. As expected, a striking difference is observed for the films as compared to the pellets. However, results show that the minimum value of the surface resistance is found for the film (#1B) with a reduced value of the critical temperature. This is not really surprising since the residual surface resistance is often an unpredictable function of the quality of the sample surface. This is especially true in the case of the novel superconducting $MgB_2$ material, where the procedure for thin film growth is still quite irreproducible. A comparison between our surface resistance measurements and data that have appeared in recent literature is rather encouraging: using the ordinary $\omega^2$ law and normalizing the data at 20 GHz, one can see that while results for the pellet lie in the range of values reported for similar samples ([49]: ~10 m$\Omega$, [50]: ~3 m$\Omega$), films show a surface dissipation significantly lower than previous reports ([20]: ~3 m$\Omega$, [31]: ~2 m$\Omega$, [51]: ~4 m$\Omega$).

The effect of ion milling on the surface impedance was also investigated: using an $Ar^+$ beam (0.5 KV, beam current 16 mA) a nominal 60 nm layer was removed from the surface of film #1B. The motivation for that is the observation that the procedure adopted until now for film synthesis favors the growth of Mg-rich surface layers, as reported by various authors [22, 51] and confirmed also by EDX (Energy Dispersive X-ray) analysis. Contrarily to what previously observed [51], there is no significant reduction in the surface resistance after ion milling: instead, a slight increase (10%) in the residual value was observed, together with a more pronounced field dependence in both $R_s$ and $X_s$ (see below). The magnetic penetration depth instead was not appreciably affected by the surface etching, as already reported in section IV.

In fig. 12 the behavior of the surface resistance as a function of temperature for pellets and films at low temperature and in the overall range (see inset) is shown. The measurements have been performed via the copper dielectrically loaded resonator in the symmetric configuration, therefore a couple of samples with nominally identical superconducting and transport properties have been used for each run. It is important to stress that what is measured here is the *average* electrodynamic response, therefore only a qualitative comparison between microwave losses in pellets and films can be done. $R_s(T)$ in the former case rapidly saturates at a large residual term decreasing T below $T_c/2$, whereas in the latter case it continuously decreases down to the lowest temperatures to a minimum value which is almost one order of magnitude smaller. The small bump in the surface resistance observed near $T_c$ has to be ascribed to a slight difference in the critical temperature values for the pair of films. The difference in the overall behavior shown by pellets and films is nevertheless a clear indication that the surface properties of $MgB_2$ in the microwave region are far from being optimized, with plenty of room for improvement increasing the quality of the samples.

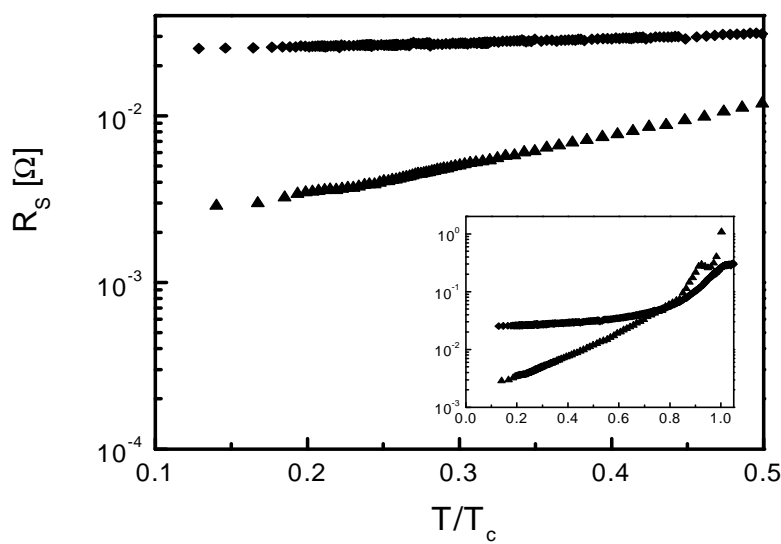

**Figure 12:** low temperature and full range (see the inset) temperature dependence of $R_s(T)$ for bulk (♦) and film (□) samples, measured using a copper dielectric resonator at 20 GHz

In figure 13 we show the temperature variation of $\lambda$ in the microwave region for $T < T_c/2$. For comparison in the same graph the low frequency penetration depth data are also displayed.

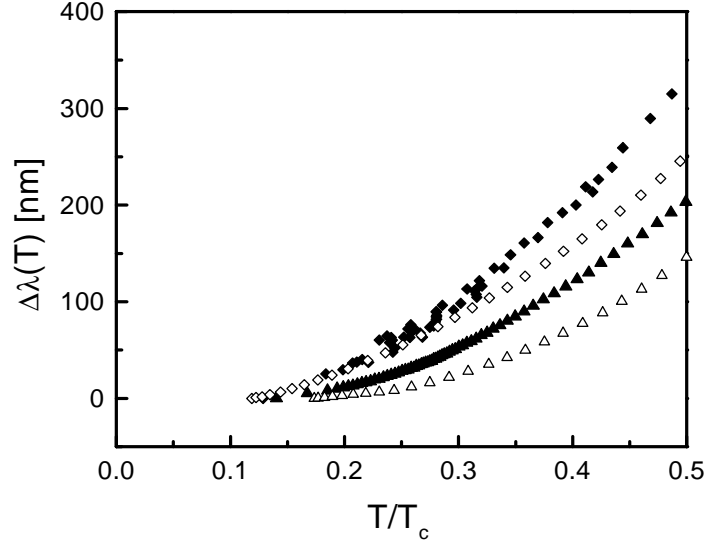

**Figure 13:** low temperature dependence of $\lambda(T)$ for bulk (♦) and film (□) samples, measured using a copper dielectric resonator at 20 GHz. For comparison the low frequency (4 MHz) magnetic penetration depth data are also shown: pellet: #2A (◊) and film #1B (△).

Pellets show a quadratic $\lambda(T)$ behavior similar to the single coil technique result and in agreement with literature data [25]. In the case of films a saturation at low temperatures, even if less pronounced, is again observed, which can be easily fitted for $T<T_c/2$ in a simple BCS framework. Using the superconducting gap and the magnetic penetration depth at zero temperature as free parameters, one yields the following result: $\Delta(0) = 2.7$ meV and $\lambda(0) = 1$ μm.

For the field measurements of the surface impedance the copper cavity was replaced with a cavity of the same geometry made of high purity niobium, and the input power varied from – 40 to + 10 dBm. Measurements were performed at liquid helium temperature.

In fig. 14 the change in the surface resistance $\Delta R_s = R_s - R_{s\ min}$, where $R_{s\ min}$ is the value at the minimum input power, versus the maximum microwave magnetic field on the sample surface, $H_{max}$, is shown for the samples under investigation.

Fig. 15 shows the change in the resonant frequency $\Delta f = f - f_{min}$, where $f_{min}$ is the value at the minimum input power, versus $H_{max}$. For small variations, the relation $\Delta f \propto \Delta \lambda$ holds, therefore this is equivalent to look at the surface reactance changes.

Together, the two dependencies provide a consistent picture of the mechanisms most likely responsible for microwave dissipation in the investigated samples. At very low fields, all samples show the same feature, characterized by a rapid increase of both the real and imaginary part of the surface impedance. In high temperature superconductors this has been commonly interpreted as flux flow losses caused by trapped flux in weak links [52]. However, this is very unlikely to happen in $MgB_2$, where magnetization [13] and non resonant microwave absorption studies [53] indicate near absence of intergranular weak links. Increasing the input power, there is a dramatic change in the dependence, since fluxoids start to enter the material. Due to the high frequency of operation, flux pinning is not encountered, and vortices can flow relatively easily. This is clearly seen in both films investigated, whereas the low values of the maximum magnetic field achieved for the pellet likely do not allow us to observe the expected variation in the $Z_s(H)$ curves.

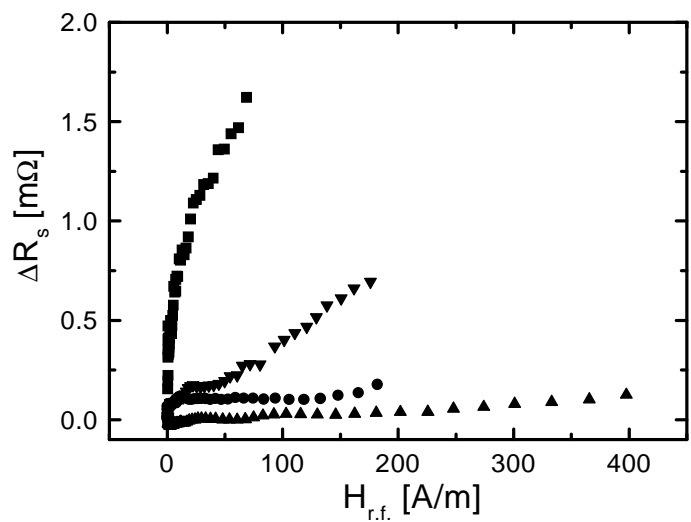

**Figure 14:** the change in the surface resistance $\Delta R_s$ as a function of the maximum microwave surface magnetic field $H_{max}$ measured at 4 K using a niobium dielectric resonator : pellet #1A (□), film #1B (□), film #1B after etching (□), film #2B (●)

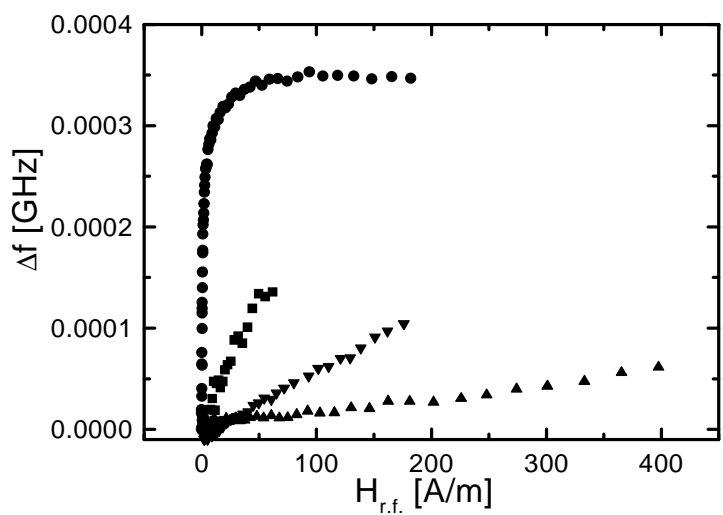

**Figure 15:** the change in the resonant frequency $\Delta f$ as a function of the maximum microwave surface magnetic field $H_{max}$ measured at 4 K using a niobium dielectric resonator: pellet #1A (□), film #1B (□), film #1B after etching (□), film #2B (●)

It is worth mentioning that, as briefly discussed above, ion milling the surface of film #1B seems to have only the effect of increasing the slope in the surface impedance dependence on field.

The most striking difference between the two films is observed in the field dependence of the surface reactance ($\Delta f \propto \Delta X_s$) in fig. 15. While this quantity displays the same dependence as $\Delta R_s$ in film #1B, its behavior is markedly different from its real counterpart in film #2. In this latter sample, non linearity in the inductance is the primary source of microwave losses. To our understanding, this is the fingerprint that the dominant mechanism beneath the microwave losses shown by the two films is not the same. This is confirmed by the analysis of the dimensionless parameter $r = \Delta R_s/\Delta X_s = \Delta(1/Q)/(\Delta f/f_{min})$, which has been widely used for different resonators to identify the source of nonlinearity in superconducting materials since it is independent of any geometrical factor [52]. In fact, the data for the three samples in fig. 16 show that the slope (i.e.: the *r* parameter), lies in a range of values between 1 and 2 for the pellet and film #1B, whereas it is one order of magnitude smaller for film #2B. These distinctive features indicate that nonlinearity in the electrodynamic response is caused by different mechanisms, likely flux flow losses in film #1 and penetration of vortices in weakly coupled grains in film #2B [54].

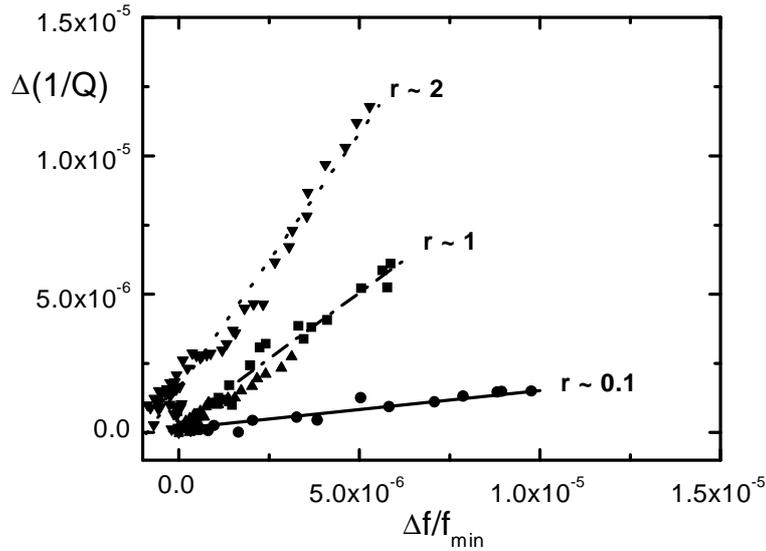

**Figure 16:** $\Delta(1/Q)$ vs $\Delta f/f_{min}$ at 4 K: pellet #1A (□), film #1B (□), film #1B after etching (□), film #2B (●). The slope of the linear fits indicates the dimensionless parameter *r* for each sample.

## VI Conclusions

The electrodynamic response of the novel intermetallic superconductor $MgB_2$ in the r.f. and microwave region has been investigated. The behavior of the magnetic penetration depth as a function of temperature and of the microwave surface impedance as a function of temperature and surface magnetic field amplitude in bulk and thin film samples has been reported. An anisotropic s-wave BCS model can account for the temperature dependence experimentally observed in the penetration depth data, confirming previous reports on the conventional nature of superconductivity in diborides. The study of the surface impedance versus temperature and field shows that the source of microwave loss can be markedly different, depending on the structural and transport properties of the samples. Pellets show very high values of losses already at $T_c/2$ and a very pronounced nonlinearity increasing the input power. Surprisingly, film #1B, which exhibits the lowest critical temperature (~ 26 K), shows also the smallest residual losses at 20 GHz. Moreover, both films retain a quite flat $R_s$ field dependence, but in film #1B field penetrates the bulk material whereas in film #2B granularity seems to be the mechanism most likely responsible for the observed microwave nonlinearity.

The difference in the overall behavior shown by pellets and films is an indication that the surface properties of $MgB_2$ in the microwave region still needs to be optimized, with plenty of room for improvement increasing the quality of the samples.

We thank F. Lombardi for her help with the ion beam etching of the film surface. Many thanks are due to A. Barone and R. Vaglio for useful discussions. We are grateful to N. Bontemps and J. Bok for fruitful discussions and for letting us to use the scientific equipment for the single coil inductive measurements. Edison acknowledges the Lecco Laboratory of the CNR-Tempe for making available its facilities for the samples preparation.

## VII References

[1] J. Namagatsu, N. Nakagawa, T. Muranaka, Y. Zenitani and J. Akimitsu, Nature (London) **410**, 63 (2001).
[2] W. L. McMillan, Phys. Rev. **167**, 331 (1968).
[3] S. L. Bud'ko, G. Lapertot, C. Petrovic, C. E. Cunningham, N. Anderson and P. C. Canfield, Phys. Rev. Lett. **86**, 1877 (2001).
[4] H. Kotegawa, K. Ishida, Y. Kitaoka, T. Muranaka and J. Akimitsu, Phys. Rev. Lett. **87**, 127001 (2001).
[5] X. K. Chen, M. J. Kostantinovic, J. C. Irwin, D. D. Lawrie and J. P. Frank, Phys. Rev. Lett. **87**, 157002 (2001).
[6] G. R. Bollinger, H. Suderow, S. Vieira, cond-mat/0102242.
[7] A. Sharoni, I. Felner and O. Millo, Phys. Rev. B **63**, 220508(R) (2001).
[8] G. Karapetrov, M. Iavarone, W. K. Kwok, G. W. Crabtree, and D. G. Hinks, Phys. Rev. Lett. **86**, 4374 (2001).
[9] A. V. Pronin, A. Pimenov, A. Loidl and S. I. Krasnosvobodtsev, Phys. Rev. Lett. **87**, 097003 (2001).
[10] S. Xu, Y. Moritomo, K. Kato and A. Nakamura, cond-mat/0104534.
[11] Y. Moritomo and S. Xu, cond-mat/0104568.

[12] A. Y. Liu, I. I. Mazin and J. Kortus, Phys. Rev. Lett. **87**, 087005 (2001).
[13] D. C Larbalestier, L. D. Cooley, M. O. Rikel, A. A. Polyanskii, J. Jang, S. Patnaik, X. Y. Cai, D. M. Fldmann, A. Gurevich, A. A. Squitieri, M. T. Naus, C. B. Eom, E. E. Hellstrom, R. J. Cava, K. A. Regan, N. Rogado, M. A. Hayward, T. He, J. S. Slusky, P. Khalifah, K. Inumaru and M. Haas , Nature (London) **410**, 186 (2001).
[14] M. Hein, *High-temperature superconductor thin films at microwave frequencies*, Springer Tracts in Modern Physics **155**, Springer-Verlag, Berlin, 1999.
[15] R. Vaglio, *Superconducting cavities*, to be published in NATO ASI Series, 2001.
[16] R. S. Gonnelli, A. Calzolari, D. Daghero, G. A. Ummarino, G. Giunchi, S. Ceresara and G. Ripamonti, Phys. Rev. Lett. **87**, 97001 (2001).
[17] Edison S.p.A. patent pending.
[18] L. Gozzellino, F. Laviano, D. Botta, A. Chiodoni, R. Gerbaldo, G. Ghigo, E. Mezzetti, G. Giunchi, S. Ceresara, G.Ripamonti, M. Poyer, cond-mat/0104069.
[19] S. X. Dou, X. L. Wang, J. Horvat, D. Milliken, A. H. Li, K. Kostantinov, E. W. Collings M. D. Sumption and H.K. Liu, Physica C **361**, 79 (2001).
[20] A. A. Zhukov, K. Yates, G. K. Perkins, Y. Bugoslavsky, M.Polichetti, A. Berenov, J. Driscoll, A. D. Caplin and L. F. Cohen, Supercond. Sci. Techn. **14**, L13 (2001).
[21] R. P. Vasquez, C. U. Jung, M.-S. Park, H.-J. Kim, J. Y. Kim, and S.-I. Lee, Phys. Rev. B **64**, 52510 (2001).
[22] M. Paranthaman, C. Cantoni, H. Y. Zhai, H. M. Christen, T. Aytug, S. Sathyamurthy, E. D. Specht, J. R. Thompson, D. H. Lowndres, H. R. Kerchner and D. K. Christen, Appl. Phys. Lett. **78**, 3669 (2001).
[23] H. M. Christen, H. Y. Zhai, C. Cantoni, M. Paranthaman, B. C. Sales, C. Rouleau, D. P. Norton, D. K. Christen, and D. H. Lowndes, Physica C **353**, 157 (2001).
[24] N. Klein, H. Chaloupka, G. Müller, S. Orbach, H. Piel, B. Roas, L. Schultz, U. Klein, and M. Peiniger, J. Appl. Phys. **67**, 6940 (1990).
[25] C. Panagopoulos, B. D. Rainford, T. Xiang, C. A: Scott, M. Kambara and I. H. Inoue, Phys. Rev. B **64**, 94514 (2001).
[26] X. H. Chen, Y. Y. Xue, R. L. Meng and C. W. Chu, cond-mat/0103029.
[27] Ch. Niedermayer, C. Bernhard, T. Holden, R. K. Kremer, and K. Ahn, cond-mat/0108431.
[28] F. Manzano and A. Carrington, cond-mat/0106166.
[29] R. Prozorov, R. W. Giannetta, S. L. Bud'ko, and P. C. Canfield, cond-mat/0108107.
[30] F. Manzano, A. Carrington, N. E. Hussey, S. Lee, and A. Yamamoto, cond-mat/0110109.
[31] N. Klein, B. B. Jin, J. Schuster, H.R. Yi, A. Pimenov, A. Loidl, S. I. Krasnovobodtsev, cond-mat/0107259.
[32] A. A. Zhukov, L. F. Cohen, A. Purnell, Y. Bugolavsky, A. Berenov, J. L. MacManus-Driscoll, H. Y. Zhai, H. M. Christen, M. P. Paranthaman, D. H. Lowndes, M. A. Jo, M. C. Blamire, L. Hao and J. Gallop, cond-mat/0107240.
[33] A. Gauzzi, J. Le Cochec, G. Lamura, B. J. Jönsson, V. A. Gasparov, F. R. Ladan, B. Plaçais, P. A. Probst, D. Pavuna and J. Bok, Rev. Sci. Instr. **71**, 2147 (2000).
[34] Z.-Y. Shen, C. Wilker, P. Pang, W. L. Holstein, D. Face, and D. J. Kountz, IEEE Trans. Microwave Theory Techn. **40**, 2424 (1992).
[35] J. Mazierska and R. Grabovickic, IEEE Trans. Applied Supercond. **8**, 178 (1998).
[36] R. A. Kaindl, M. A. Carnahan, J. Orenstein, D. S. Chemla, H. M. Christen, H. Zhai, M. Paranthaman, D. H. Lowndes, cond-mat/0106342.

[37] L. Cohen, private communication.
[38] A. M. Gabovich and A. I. Voitenko, Phys. Rev. B, **60**, 7465 (1999).
[39] A. L. Schawlow and G. E. Devlin, Phys. Rev. **113**, 120 (1959).
[40] F. Giubileo, D. Roditchev, W. Sacks, R. Lamy and J. Klein, cond-mat/0105146.
[41] F. Giubileo, D. Roditchev, W. Sacks, R. Lamy, D. X. Thanh, J. Klein, S. Miraglia, D. Fruchart, J. Markus and Ph. Monod, cond-mat/0105592.
[42] N. Klein, N. Tellmann, H. Schulz, K. Urban, S. A. Wolf and V. Z. Kresin , Phys. Rev. Lett. **71**, 3355 (1993).
[43] S. Tsuda, T. Toyoka, T. Kiss, Y. Takano, K. Togano, H. Kitou, H. Ihara and S. Shin, cond-mat/0104489.
[44] P. Szabó, P. Samuely, J. Kacmarcík, T. Klein, J. Marcus, D. Fruchart, S. Miraglia, C. Marcenat and A. G. M. Jansen, Phys. Rev. Lett. **87**, 137005 (2001).
[45] F. Laube, G. Goll, J. Hagel, H. V. Lohneysen, D. Ernstand Wolf, cond-mat/0106407.
[46] Y. Wang, T. Plackowski and A.Junod, Physica C **355**, 179 (2001).
[47] A. A. Abrikosov and L. P. Gor'kov, Soviet Phys. JETP **12**, 1243(1961).
[48] M. Xu, H. Kitazawa, Y. Takano, J. Ye, K. Nischida, H. Abe, A. Matsushita and G. Kido, cond-mat/0105271.
[49] N. Hakim, P. V. Parimi, C. Kusko, S. Sridhar, P. C. Canfield, S. L. Bud'ko, and D. K. Finnemore, Appl. Phys. Lett. **78**, 4160 (2001).
[50] Yu. A. Nefyodov, M. R. Trunin, A. F. Shevchun, D. V. Shovkun, N. N. Kolesnikov, M. P. Kulakov, A. Agliolo Gallitto, and S. Fricano, cond-mat/0107057.
[51] S. Y. Lee, J. H. Lee, J. S. Ryu, J. Lim, S. H. Moon, H. N. Lee, H. G. Kim, and B. Ohh, cond-mat/0105327.
[52] J. Halbritter, Phys. Rev. B **48**, 9735 (1993).
[53] J. P. Joshi, S. Sarangi, A. K. Sood, S. V. Bhat, and D. Pal, cond-mat/0103369.
[54] M. A. Golosovsky, H. J. Snortland, and M. R. Beasley, Phys. Rev. B **51**, 6462 (1995)